\documentstyle[12pt,aasms4]{article}

\newcommand{\ie}{i.e.}
\newcommand{\eg}{e.g.}
\newcommand{\etal}{et al.}
\newcommand{\f}{F}
\newcommand{\fo}{{\rm FO}}
\newcommand{\so}{{\rm SO}}
\newcommand{\PL}{P$-$L}
\newcommand{\Teff}{${\rm T}_{\rm eff}$}
\newcommand{\socan}{MACHO*05:03:39.6$-$70:04:32}
\newcommand{\vanicek}{Van\'{\i}\v{c}ek}

\lefthead{Alcock \etal}
\righthead{\so\ Pulsation Mode from \fo/\so\ Beat Cepheids}

\begin{document}

\title{The MACHO Project LMC Variable Star Inventory. VI. The 
Second-overtone Mode of Cepheid Pulsation From First/Second Overtone 
(\fo/\so) Beat Cepheids}

\author{C.~Alcock\altaffilmark{1,2}, 
      R.A.~Allsman\altaffilmark{3},
        D.~Alves\altaffilmark{1,4},
      T.S.~Axelrod\altaffilmark{5},
      A.C.~Becker\altaffilmark{6},
      D.P.~Bennett\altaffilmark{1,2,7},
      K.H.~Cook\altaffilmark{1,2},
      K.C.~Freeman\altaffilmark{5},
        K.~Griest\altaffilmark{2,8},
      M.J.~Lehner\altaffilmark{2,8,9},
      S.L.~Marshall\altaffilmark{1,2},
        D.~Minniti\altaffilmark{1,2},    
      B.A.~Peterson\altaffilmark{5},
      M.R.~Pratt\altaffilmark{2,6,10},
      P.J.~Quinn\altaffilmark{11},
      A.W.~Rodgers\altaffilmark{5,14},
        A.~Rorabeck\altaffilmark{12},
        W.~Sutherland\altaffilmark{13},
        A.~Tomaney\altaffilmark{6},
        T.~Vandehei\altaffilmark{2,8},
      D.L.~Welch\altaffilmark{12}\\
      {\bf (The MACHO Collaboration)}
}

\altaffiltext{1}{Lawrence Livermore National Laboratory, Livermore, CA 94550
        E-mail: {\tt alcock, alves, dminniti, kcook, stuart@igpp.llnl.gov}}
 
\altaffiltext{2}{Center for Particle Astrophysics,
        University of California, Berkeley, CA 94720}
 
\altaffiltext{3}{Supercomputing Facility, Australian National University,
        Canberra, ACT 0200, Australia
        E-mail: {\tt robyn@macho.anu.edu.au}}
 
\altaffiltext{4}{Department of Physics, University of California,
        Davis, CA 95616 }
 
\altaffiltext{5}{Mt.~Stromlo and Siding Spring Observatories, Australian
        National University, Weston Creek, ACT 2611, Australia
        E-mail: {\tt tsa, kcf, peterson@mso.anu.edu.au}}
 
\altaffiltext{6}{Departments of Astronomy and Physics,
        University of Washington, Seattle, WA 98195
        E-mail: {\tt becker, austin@astro.washington.edu}}

\altaffiltext{7}{Physics Department, University of Notre Dame, Notre 
        Dame, IN 46556
        E-mail: {\tt bennett.27@nd.edu}}
 
\altaffiltext{8}{Department of Physics, University of California,
        San Diego, La Jolla, CA 92093
        E-mail: {\tt kgriest, tvandehei@ucsd.edu }}

\altaffiltext{9}{Department of Physics, University of Sheffield,
        Sheffield, S3 7RH, U.K.
        E-mail: {\tt M.Lehner@sheffield.ac.uk}}

\altaffiltext{10}{LIGO Project, MIT, Room 20B-145, Cambridge, MA 02139
        E-mail: {\tt mrp@ligo.mit.edu}}
 
\altaffiltext{11}{European Southern Observatory, Karl-Schwarzchild Str. 2,
        D-85748, Garching, Germany
        E-mail: {\tt pjq@eso.org}}

\altaffiltext{12}{Dept. of Physics \& Astronomy, McMaster University,
        Hamilton, Ontario, L8S 4M1 Canada
        E-mail: {\tt welch@physics.mcmaster.ca, rorabeck@glen-net.ca}}

\altaffiltext{13}{Department of Physics, University of Oxford, Oxford OX1
        3RH, U.K.
        E-mail: {\tt w.sutherland@physics.ox.ac.uk}}

\altaffiltext{14}{Deceased.}

\setcounter{footnote}{0}

\begin{abstract}

MACHO Project photometry of 45 LMC \fo/\so\ beat Cepheids which
pulsate in the first and second overtone (\fo\ and \so, respectively) 
has been analysed to determine the lightcurve characteristics for the  
\so\ mode of Cepheid pulsation.  We predict that singly-periodic \so\ 
Cepheids will have nearly sinusoidal lightcurves; that we will only be 
able to discern \so\ Cepheids from fundamental (F) and (\fo) Cepheids 
for $P \lesssim 1.4$ days;  and that the \so\ distribution will overlap 
the short-period edge of the LMC \fo\ Cepheid period-luminosity relation 
(when both are plotted as a function of photometric period).

We also report the discovery of one \so\ Cepheid candidate, \socan, with a
photometric period of $0.775961\pm 0.000019$ days and an instrumental
amplitude of $0.047\pm 0.009$ mag in V. 

\end{abstract}


\section{Introduction}

Gravitational microlensing surveys have produced photometric databases of
unprecedented usefulness --- both in the number of stars monitored and in the
quality and length of their time series.  Such databases are favorable
hunting grounds for rare (and not-so-rare) types of variable stars and
stellar systems. 

Among the brightest and most important variable stars are the classical
Cepheids.  These stars are arguably the most thoroughly understood
intrinsic variable from a theoretical standpoint, and are of paramount 
importance for calibrating the extragalactic distance scale.  Microlensing 
surveys have expanded the number of known Cepheids, which have served to clarify
our understanding of Cepheid pulsation. In this paper, we investigate the 
second-overtone (\so) mode of Cepheid pulsation using LMC first/second overtone 
(\fo/\so) beat Cepheids discovered by the MACHO Project. 

\section{Theoretical Motivation}

Stobie \markcite{Sto-69a}(1969a) reported the first theoretical findings
suggesting that Cepheids might be unstable in the \so\ (radial) mode. 
His non-linear, radiative calculations predicted the existence of \so\
Cepheids for low \L\ and high \Teff, which lead Stobie
\markcite{Sto-69b}(1969b) to suggest that such Cepheids might comprise the
short-period peak in the SMC Cepheid period distribution.  While some
early investigations had difficulty in producing \so\ Cepheids (\eg, Cox
\markcite{Cox-80}1980 mentioned the \so\ mode ``is frequently not even
pulsationally unstable in the linear theory''), most nonlinear
calculations (\eg, Chiosi, Wood \& Capitano \markcite{CWC-93}1993) have
confirmed the positive growth rate of the \so\ mode for Cepheid surface
gravities and temperatures, thus predicting that \so\ Cepheids should
exist, provided that real stars could reach the region of instability
by normal evolution.

\section{Observational Motivation}

Observational evidence of \so\ mode excitation in singly-periodic Cepheids
has lagged theory by more than two decades.  It was only in the 1980s that
mode identification of radial pulsators became observationally feasible,
guided by predicted period ratios from linear and non-linear calculations:
\ie, $P_\fo/P_\f\simeq 0.7$ for the first-overtone (\fo) and fundamental
(\f) modes, and $P_\so/P_\fo\simeq 0.8$.  Two methods are widely used to
identify in which modes a Cepheid pulsates: Fourier decomposition of
lightcurves, which has become the standard for comparing theory and
observation (\eg, Moskalik, Buchler \& Marom \markcite{MBM-92}1992); and
position on the observed \PL\ relation for extragalactic Cepheids 
(the \f\ \PL\ relation or the \fo\ \PL\ relation, as delimited by
B\"{o}hm-Vitense \markcite{BV-94}1994 or Alcock \etal\ 
\markcite{Beat1-95}1995). 

To date, there exists one possible singly-periodic \so\ Cepheid candidate:
HR 7308, a Galactic Population I Cepheid with constant period (1.491
days), but varying radial velocity (2.3$-$35 km s$^{-1}$) and V
semi-amplitude (0.06$-$0.17 mag; Andrievsky, Kovtyukh \& Usenko
\markcite{AKU-94}1994).  Although HR 7308 is consistent with some \so\
Cepheid models (Burki \etal\ \markcite{Bur-86}1986; Fabregat, Suso \&
Reglero \markcite{FSR-90}1990;  Simon \markcite{Sim-85}1985), there are
observations which identify it as a \fo\ or \f\ pulsator (Bersier
\markcite{Ber-96}1996;  Bersier \& Burki \markcite{BB-96}1996).  Thus, it
is fair to say that no singly-periodic \so\ Cepheids have yet been
conclusively identified. 

Unlike singly-periodic pulsators, beat Cepheids act as calibrators
for mode identification: because they pulsate in two modes simultaneously,
their period ratios identify their modes of pulsation.  Prior to the
1980s, only \f/\fo\ beat Cepheids were known to exist. CO Aur, the
first (and, so far, only) Galactic \fo/\so\ beat Cepheid, was discovered
and analysed by Mantegazza \markcite{Man-83}(1983) and Antonello \& Mantegazza
\markcite{AM-84}(1984).  However, one star constituted too small a sample 
for extracting the characteristics of \so\ Cepheid pulsators. This situation 
changed when Alcock \etal\ \markcite{Beat1-95}(1995) provided the first 
conclusive observational evidence of \so\ mode excitation by reporting 15
LMC \fo/\so\ beat Cepheids.  Alcock \etal\ also suggested that
singly-periodic \so\ Cepheids might be found in the MACHO Project LMC
database.  The motivation for discovering \so\ Cepheids is
straightforward:  the common distance of such stars in the LMC, as well as
its small amount of differential reddening should allow us to determine 
unambiguously the $\log{\rm L}-\log{\rm T}_{\rm eff}$ region where \so\ 
excitation occurs.

In subsequent sections we take the first steps towards pursuing this goal,
making use of the results of Fourier decomposition of LMC \fo/\so\ beat
Cepheid lightcurves to estimate the Fourier parameters of singly-periodic
\so\ pulsators. 

\section{Fourier Decomposition of Lightcurves: Method}

We utilized an expanded sample (compared to Alcock \etal\
\markcite{Beat1-95}1995; Welch \etal\ \markcite{Wel-97}1997) of 47
confirmed \fo/\so\ beat Cepheid lightcurves in the LMC discovered by the
MACHO Project.  Only 45 beat Cepheids are unique: there are two beat Cepheids
which were observed in two overlapping MACHO fields. 

We refer the reader to Alcock \etal\ \markcite{Beat1-95}(1995) for a
description of our observations and the beat Cepheid identification
process.  Briefly, each star has two-bandpass photometry (a `V' and `R'
band) taken over four years of observation, which results in a time-series
for each star of 600--1000 observation epochs.  Automatic reduction and
analysis routines for these lightcurves provide quality flags for each
observation.  We used these flags to remove any observations which were
suspect because of possible cosmic ray events, bad or missing pixels, or
poor image quality.  This typically resulted in the rejection of up to
30\% of the observations for a given star, since some of these stars are
relatively faint. 

Each beat Cepheid in our sample can be characterized as pulsating with two
non-commensurate frequencies, $\nu_\fo$ and $\nu_\so$, and their linear
combinations: \ie, any frequencies $\nu=i\nu_\fo+j\nu_\so >0$, with $i$
and $j$ integer.  These frequencies are attributed to \fo\ and \so\
interaction within the star. Our goal is go isolate the \so\ mode of
pulsation for each beat Cepheid through Fourier decomposition: \ie, we
model a given beat Cepheid's lightcurve by a truncated Fourier series,
examining only the model terms that belong to solely the (\fo\ and) \so\
modes, ignoring any mixing terms (which typically have much smaller
amplitudes than he principle terms).  As Pardo \& Poretti
\markcite{ParP-97}(1997) pointed out, traditional Fourier analysis
involved modeling lightcurves by truncated Fourier series of an \emph{a
priori} order, without testing to see which model terms were warranted. 
To avoid this concern, we opted to code the comprehensive CLEANest
algorithm of Foster (\markcite{Fos-95}1995;  \markcite{Fos-96a}1996a;
\markcite{Fos-96b}1996b) for joint frequency analysis and lightcurve
modeling.  Foster's algorithm utilizes an accurate power spectrum
estimator (the date-compensated discrete Fourier transform (DCDFT) of
Ferraz-Mello \markcite{FM-81}1981) and provides both frequency and model
parameter uncertainties.  As well, Foster (\markcite{Fos-96a}1996a;
\markcite{Fos-96b}1996b) provides statistical tests of significance for
frequencies in a spectrum, and a thorough discussion of how CLEANest
differs from other well-known spectral analysis routines. 

Our analysis of each beat Cepheid resembles that of Pardo \& Poretti
\markcite{ParP-97}(1997).  On the first application to a lightcurve, our
coding of CLEANest uses the DCDFT of the photometry to produce a power
spectrum in the frequency range $\nu_{\rm res}=(2T_{\rm span})^{-1}$ to
$\nu_{\rm max}=(2{\rm min}(\Delta t))^{-1}$ in stepsizes of $\nu_{\rm
res}$ (the frequency resolution).  $T_{\rm span}$ is the total time span
of the observations for a star, and $\Delta t$ the minimum time separation 
between successive observations.  If any frequencies in the DCDFT are adopted as
significant (the criteria for selection are presented below), each frequency
$\nu_i$ is modeled by $\cos(2\pi\nu_i t)$ and $\sin(2\pi\nu_i t)$ terms,
plus an overall constant, as prescribed in Foster \markcite{Fos-95}(1995). 
The resulting model is subtracted from the original photometry, forming
residuals; these residuals are subjected to another DCDFT, and the process
is iterated until no significant periodic variations remain in the
residuals.  Each time a DCDFT of the data or residuals has been performed,
CLEANest seeks to find the $n$-tuple of frequencies which gives the best
description of the data, by varying each frequency in its neighbourhood
until a maximum of Foster (\markcite{Fos-96a}1996a;
\markcite{Fos-96b}1996b)'s model amplitude is found.  We allowed CLEANest
to perform this frequency variation without requiring that model
frequencies retained their expected relations to $\nu_\fo$ and $\nu_\so$;
the fact that most model frequencies did retain their expected relation to
$\nu_\fo$ and $\nu_\so$ attested to their authenticity. 

What determines whether or not a frequency is significant? $\nu_\fo$ and
$\nu_\so$ were usually found to be the two most powerful frequencies in
the first couple DCDFTs of the data\footnote{$\nu_\fo$ and $\nu_\so$ were
confirmed by the expectation that $\nu_\fo/\nu_\so\simeq 0.805$ (Alcock
\etal\ \markcite{Beat1-95}1995).}, and usually passed Foster
\markcite{Fos-96a}(1996a)'s test for statistical significance. 
Unfortunately, most linear combinations of $\nu_\fo$ and $\nu_\so$ that we
found in DCDFTs failed to pass tests of significance. We thus adopted a
frequency as significant if it was a linear combination of $\nu_\fo$ and
$\nu_\so$ (within estimated uncertainties);  if it appeared as one of the
20 most powerful frequencies in a residual DCDFT; and if it was reasonable
(\eg, we would not have modeled a frequency that seemed to be
$2\nu_\fo+\nu_\so$ if we had not previously detected $2\nu_\fo$ or
$\nu_\so$ in our analysis).  We terminated our modeling when no
frequencies consistent with a linear combination of $\nu_\fo$ and
$\nu_\so$ could be identified in the 20 most powerful frequencies in a
residual DCDFT.  In most cases, we also had to adopt a model term with a
frequency close to 1.003 day$^{-1}$, corresponding to one sidereal day, to
make apparent further linear combinations of $\nu_\fo$ and $\nu_\so$.  In
a few cases, we also adopted a model term with one very low frequency,
typically around 0.0005 day$^{-1}$, in order to permit further analysis. 
Scheduling of observations were responsible for the appearance of the
sidereal frequency.  The small frequency could arise from a zero-point
shift in observations, or from a long-period amplitude evolution of the
zero-point (mean) of a star, or perhaps from some other cause.  Whatever
its genesis in a given star, a small frequency has no appreciable effect
on our analysis or conclusions, and we do not discuss these further. 

In order to test for robustness of CLEANest's model parameters, we
subjected all model parameters in our final CLEANest models to a Marquardt
fitting algorithm, a $\chi^2$ minimization method which pragmatically
alternates between a `steepest descent' (or gradient-search) algorithm
when $\chi^2$ changes rapidly near a given set of model parameters, and a
first-order model parameter expansion when $\chi^2$ changes little near a
given set of model parameters (\eg, Bevington \& Robinson
\markcite{BR-92}1992).  Most model coefficients and frequencies remained
within the uncertainties of their CLEANest values, and retained their
relationship to $\nu_\fo$ and $\nu_\so$ within uncertainties. 

In Table \ref{pertable} we present identifiers, periods, and period ratios
for our \fo/\so\ stars.  There are two more \fo/\so\ stars than tabulated
in Welch \etal\ \markcite{Wel-97}(1997).  While our results here are not
as precise as those listed by Welch \etal, they supercede those values.
The uncertainties derived in this study are also believed to be 
better estimates. 

Finally, Figure \ref{example-fit} illustrates the results of our fit
procedure for a typical \fo/\so\ beat Cepheid.  We show lightcurves of the
composite, \fo, and \so\ modes of pulsation, along with frequency
information.  The \fo\ and \so\ mode lightcurves were derived by
prewhitening observations with any model terms not belonging to the
displayed mode, as well as mixing terms between modes.  The lightcurve
shown had the magnitude-average of its observations removed, and is on the
instrumental V-band of the MACHO Project.  \placefigure{example-fit}

\section{Fourier Parameters for the \so\ Mode of Cepheid Pulsation}

Our analysis yielded 37 beat Cepheids with robust $\nu_\fo$, $\nu_\so$,
$i\nu_\fo+\nu_\so$, and $i\nu_\fo$ frequencies (with $i$ up to 4), as well
as 8 stars with the above plus the $2\nu_\so$ harmonic.  In no case did we
detect harmonics higher than $2\nu_\so$ for the \so\ mode.  We used model
coefficients from the \fo\ and \so\ model terms to form the usual
lightcurve geometry indicators: $R_{k1}$, the amplitude ratio of the
$(k-1)$th harmonic and the `base' frequency model terms for a mode; and
$\phi_{k1}=\phi_k-k\phi_1$, where $\phi_k$ is the phase of a mode's
$(k-1)$th harmonic, and $\phi_1$ the phase of its `base' frequency model
term\footnote{Original use of $R_{k1}$ and $\phi_{k1}$ is commonly
attributed to Simon \& Lee \markcite{SL-81}(1981), but related quantities
were used and defined in \eg, Payne-Gaposchkin \markcite{PG-47}(1947) and
Kukarkin \& Parenago \markcite{KP-37}(1937).  Note that $\phi_{k1}$ is
independent of reference epoch.}. The absence of any stars with \so\
harmonics higher than $2\nu_\so$ clearly limits us to $R_{21}$ and
$\phi_{21}$ for descriptions of the \so\ mode's lightcurve shape, even
though up to $R_{41}$ and $\phi_{41}$ were possible for the \fo\ mode.  We
will present these higher-order Fourier decomposition parameters for the
\fo\ mode in a future paper. 

The Fourier parameters $R_{21}$ and $\phi_{21}$ for the \fo\ and \so\
modes are shown in Figures \ref{R21} and \ref{P21}, plotted against the
\fo\ and \so\ mode periods, respectively: \ie, as their singly-periodic
counterparts would appear in such plots.  All stars appear in the \fo\
sequences. 

Only a small fraction of our beat Cepheids clearly exhibited $2\nu_\so$,
needed to form $R_{21}$ and $\phi_{21}$.  In order to try and draw out as
much information on the \so\ mode as possible, we adopted a $2\nu_\so$
from the best-fit $\nu_\so$ for each of the 37 stars which did not exhibit
a $2\nu_\so$ variation using CLEANest, and attempted to fit model terms for
such a frequency with our Marquardt algorithm, keeping only those stars
whose best-fit result was consistent with our initial guess.  Eighteen
extra stars had such `stable' $2\nu_\so$ frequencies.  We added them to
Figures \ref{R21} and \ref{P21}, broadening our \so\ mode sequences to a
greater range of periods ($0.49<$P$<1.1$ days).  The remaining 19 stars
failed to have a stable $2\nu_\so$ and so have a \so\ mode with a purely
sinusoidal lightcurve shape: \ie, $R_{21}=0$.  These 19 stars are shown in
Figure \ref{R21}, but not in Figure \ref{P21}, as their $\phi_{21}$ values
are undefined. \placefigure{R21} \placefigure{P21}

\section{Discussion}

\subsection{The Long-Period Limit for Identification of LMC \so\ Cepheids}

Recent theoretical investigations examining the \so\ mode of Cepheid
pulsation suggest that \so\ Cepheids can have periods of 10 days or more
(\eg, the comprehensive study of Chiosi \etal\ \markcite{CWC-96}1996).  We
are in the position to add some observational constraints on this
long-period limit. 

Figure \ref{R21-P21-for-LMC-Cephs} shows $R_{21}$ and $\phi_{21}$ for all
known LMC singly-periodic Cepheids with $R_{21}$ uncertainties less than
0.05 and $\phi_{21}$ uncertainties less than 0.1 rad, discovered by the
MACHO Project (adapted from Welch \etal\ \markcite{Wel-97}1997).  Over
1400 Cepheids are displayed, but all avoid combinations of $R_{21}$ and
$\phi_{21}$ in regions we expect for a \so\ pulsator (the hatched regions
in the Figure, derived from our Figures \ref{R21} and \ref{P21}).  If
Figure \ref{R21-P21-for-LMC-Cephs}'s sample of Cepheids is indicative of
LMC Cepheids as a whole over their period range, we would expect to see at
least some stars in the longer-period \so\ sequence range --- say around 1
day --- but we do not see any.  This is true even when restrictions on
$R_{21}$'s and $\phi_{21}$'s uncertainties are removed: no \so\ candidates
appear in the shaded region of the $R_{21}$--P portion of the Figure. 
Those stars which lie in the shaded region of the $\phi_{21}$--P portion
of Figure \ref{R21-P21-for-LMC-Cephs} have $R_{21}$ values consistent with
\fo\ and \f\ pulsators. \placefigure{R21-P21-for-LMC-Cephs}

The lack of P $\lesssim$ 1 day Cepheid candidates may be the result of a
bias --- in that Figure \ref{R21-P21-for-LMC-Cephs}'s Cepheids were
selected for periods of about 1 day or longer, as well as CMD position,
P--L position, and lightcurve shape.  However, the lack of an obvious
third Fourier parameter sequence at periods longer than 1 day is probably
real, as one would surmise from the greater density of stars in Figure
\ref{R21-P21-for-LMC-Cephs} to longer periods.  This suggests that \so\
Cepheids are indeed a rare species. 

It would be intriguing if the $R_{21}$ \so\ sequence mimicked the `V'
shape of the \fo\ and \f\ mode $R_{21}$ sequences of Figure
\ref{R21-P21-for-LMC-Cephs}.  What we see in Figure \ref{R21} would be the
short-period commencement of such a sequence, and we would expect,
analogous to the sequences in Figure \ref{R21-P21-for-LMC-Cephs}, a drop
to zero followed by a sharp rise in $R_{21}$ on the long-period side of
this drop.  Such behaviour has been predicted for \so\ Cepheids,
attributed to a resonance between the second and sixth overtone modes
around \so\ periods of 1 day (Antonello \& Kanbur \markcite{AK-97}1997). 
Presumably, the rising branch of $R_{21}$ for the \so\ mode would be
hidden in the \fo\ sequence we see in Figure \ref{R21-P21-for-LMC-Cephs}. 
This would limit our ability to discern \so\ Cepheids based on lightcurve
shape to stars with periods less than about 1.4 days. 

\subsection{Lightcurve Shape: Observations vs. Theoretical Suggestions}

The first investigation of lightcurve shape for \so\ radial pulsators by
Stellingwerf, Gautschy \& Dickens \markcite{SGD-87}(1987) suggested such
stars would have the characteristic `sawtooth' nature of \f\ lightcurves,
based on Fourier decompositions of one-zone models.  While this suggestion
was used to identify \so\ RR Lyrae stars (the `RRe' stars of Alcock \etal\
\markcite{RRe-96}1996), the Fourier parameters of Figures \ref{R21} and
\ref{P21} do not comply with such a suggestion for \so\ Cepheids, which
would require $R_{21}\gtrsim 0.3$ and $\phi_{21}\simeq 4.0$ rad.  This
non-conformity is not unexpected: as our referee noted, one-zone models of
RR Lyrae-type pulsations may not have much bearing on actual lightcurves
of \so\ Cepheids.  Antonello \& Kanbur \markcite{AK-97}(1997) have
recently completed a \so\ Cepheid pulsation study which indicates
lightcurves of a type ``more sinusoidal than [\fo] ones, at least for
short periods'' (about 1 day or less). Observationally, we find sinusoidal
--- or nearly sinusoidal --- lightcurves to be the norm for the Cepheid
\so\ mode.  Pardo \& Poretti \markcite{ParP-97}(1997) suggested as much
for CO Aur, the sole Galactic \fo/\so\ beat Cepheid, when they failed to
detect its $2\nu_\so$ frequency.  Future observational and theoretical
investigations should be guided by this finding: that \so\ Cepheids have a
symmetric, nearly-sinusoidal light variation over their pulsational cycle. 

This has a bearing on nomenclature.  Some authors use the term
`s--Cepheid' (\eg, Antonello, Poretti, \& Reduzzi \markcite{Ant-90}1990; 
Morgan \markcite{Mor-95}1995) for stars that are, largely, \fo\ Cepheids
(as shown most recently for Galactic Cepheids by Pardo \& Poretti
\markcite{ParP-97}1997).  We have shown that \so\ Cepheids are distinctly
more sinusoidal in lightcurve shape than \fo\ and \f\ Cepheids from their
low $R_{21}$ values.  It might be better to refer to s--Cepheids instead
as first overtone (\fo) Cepheids in the future, to avoid ambiguities. 

\section{A LMC \so\ Cepheid Candidate}

During the extraction and inspection of variables from a section of the
CMD for our four years of LMC observation, we discovered \socan, a
small-amplitude sinusoidal pulsator.  In Figures \ref{PL} and \ref{CMD} we
show its position in the P--L relation and CMD for known LMC Cepheids. 
Although it occupies sparsely populated regions in these diagrams, it is,
most probably, a classical Cepheid. \placefigure{PL} \placefigure{CMD}

We transformed this star's MACHO V- and R-band photometry to V and 
R$_{\rm KC}$ using our latest transformation equations, and subjected its 
observations to the fitting procedure outlined above.  The results for 
the V band are displayed as Figure \ref{V-fit}.  We found all bandpasses 
were best described by a single component,
\[
X(t)=X_0+\Delta_X\cos(2\pi(t-t_0)\nu_X+\phi_X),
\]
where $X$ is one of V or R$_{\rm KC}$, and $t_0=$ JD 2448628.632.  The 
coefficients from these fits are presented in Table \ref{coeffs}.  A 
frequency corresponding to 1.003 days$^{-1}$ was also adopted in both fits, 
but made no difference to their appearance.  \socan\ stands as a 
(serendipitous) first extragalactic \so\ Cepheid candidate.
\placefigure{V-fit}

\section{Future Directions}

We are, at present, actively searching for more singly-periodic \so\
Cepheids in the MACHO LMC and SMC databases.  Any additional \so\ candidates 
will be reported in a future paper. 

The number of short-period pulsators will increase as one moves to lower
metallicity environments (\eg, Lipunova \markcite{Lip-92}1992). Thus, the
SMC should provide an even better hunting ground for beat Cepheids and
\so\ candidates when the MACHO Project analysis of SMC photometry is
complete.

\acknowledgements

We are grateful for the skilled support given our project by the technical
staff at Mt. Stromlo Observatory (MSO). Work performed at Lawrence
Livermore National Laboratory (LLNL) is supported by the Department of
Energy (DOE) under contract W7405-ENG-48. Work performed by the Center for
Particle Astrophysics (CfPA) on the University of California campuses is
supported in part by the Office of Science and Technology Centers of the
National Science Foundation (NSF) under cooperative agreement AST-8809616. 
Work performed at MSO is supported by the Bilateral Science and Technology
Program of the Australian Department of Industry, Technology and Regional
Development. KG acknowledges a DOE OJI grant, and the support of the Sloan
Foundation. DLW and AJR were supported, in part, by a Research Grant from
the Natural Sciences and Engineering Research Council of Canada (NSERC)
during this work.  AJR was also supported, in part, by an NSERC
Postgraduate scholarship (PGS A).  This work comprised part of his M.Sc. 
thesis.

\newpage
\figcaption[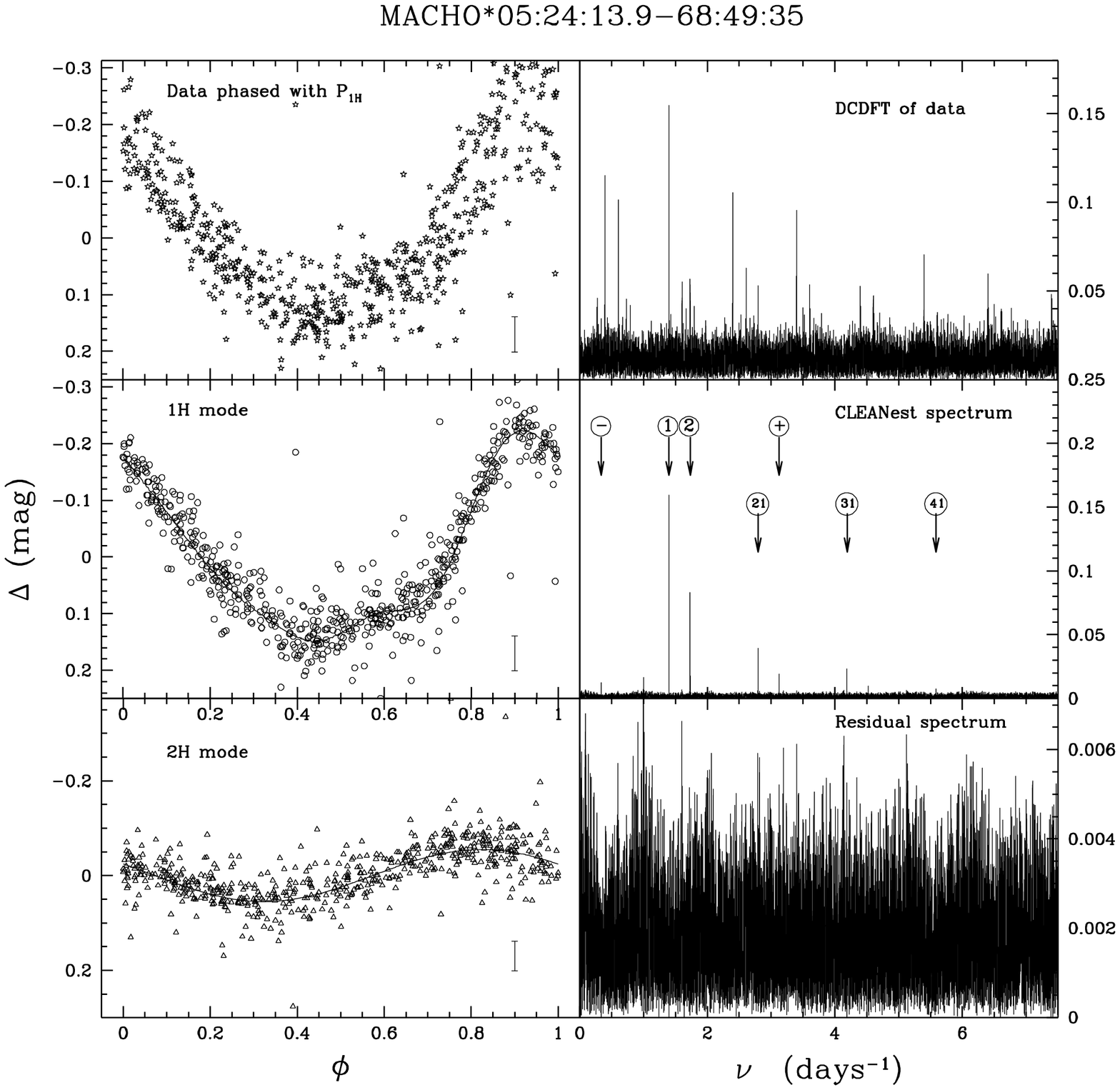]{
Model results for a typical beat Cepheid.  The top left panel
displays the lightcurve phased with the \fo\ mode period, 0.715871(18)
days.  The middle left panel displays the best-fit \fo\ mode solution,
while the bottom left panel displays the best-fit \so\ mode solution.  The
error bar on the left-hand plots is typical.  The topmost right panel
displays the amplitude spectrum (DCDFT) of the original photometry --- a
series of significant frequencies and their aliases.  The middle right
panel displays the amplitude spectrum for the DCDFT of the final
residuals, superimposed with the best-fit $n$-tuple of frequencies for the
star. The bottom right panel shows the residual spectrum of the middle
right panel on a finer scale, with the best-fit $n$-tuple frequencies
removed.  No significant frequencies are apparent.\label{example-fit}} 

\figcaption[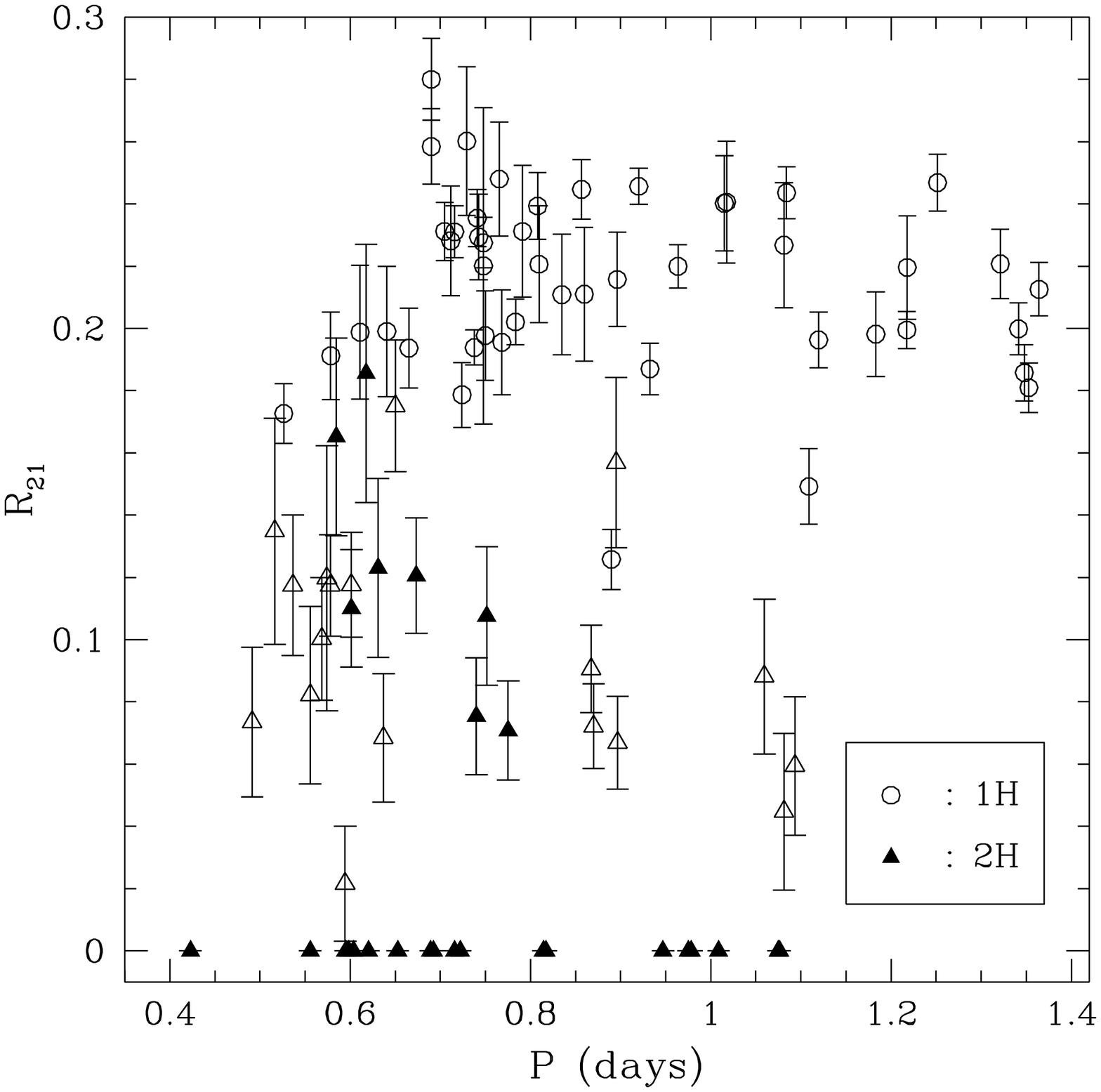]{
\label{R21}
The amplitude ratio $R_{21}$ for the \fo\ and \so\ modes of our 
beat Cepheids, using all stars with stable $2\nu_\so$ frequencies, 
whether they were detected using the CLEANest algorithm (8 filled 
triangles) or introduced after using CLEANest, but found to be stable (18 
open triangles).  The remaining stars with $R_{21}=0$ are also shown.}

\figcaption[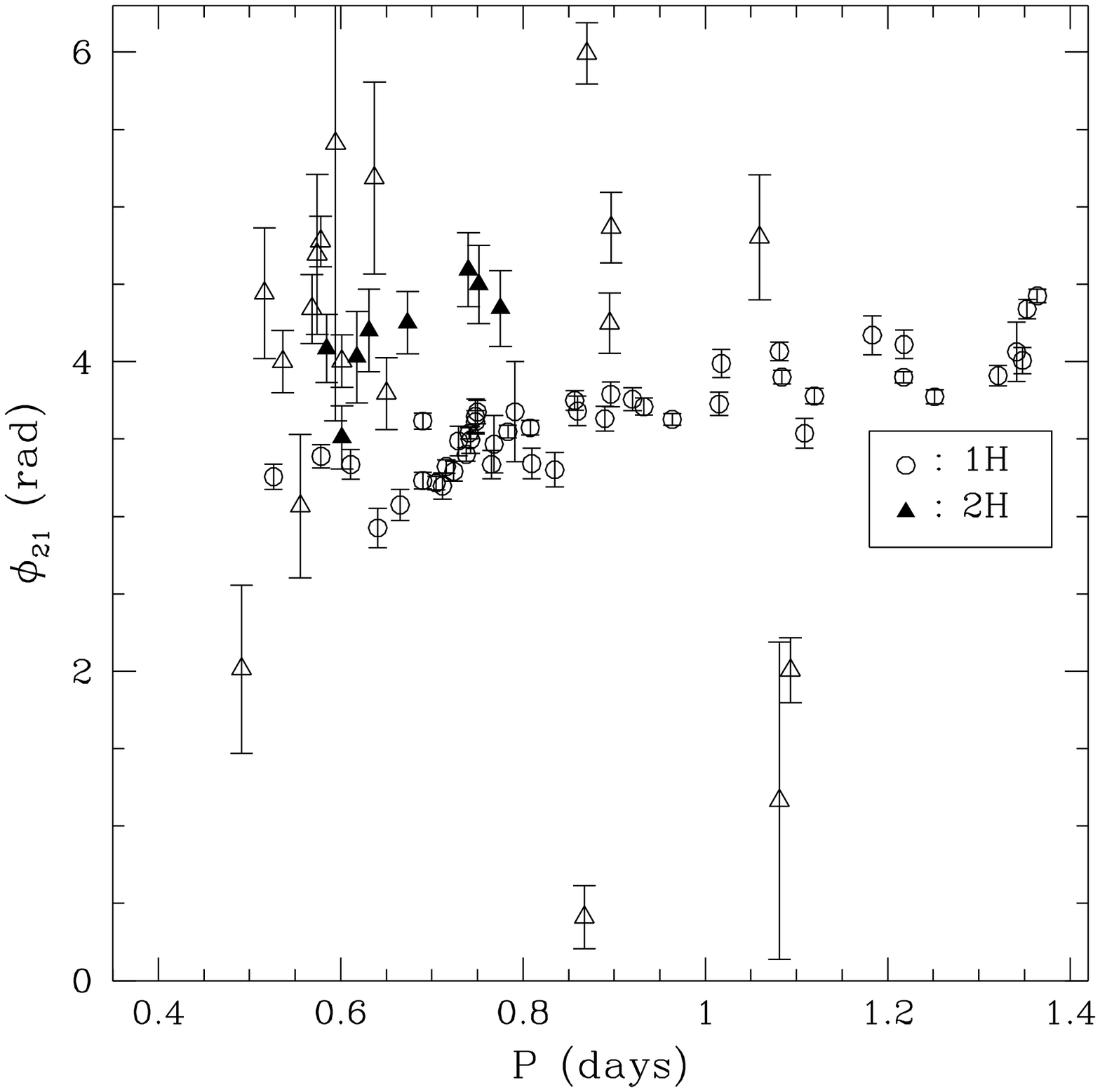]{
\label{P21}
The Fourier phase $\phi_{21}$ for the \fo\ and \so\ modes of our
beat Cepheids, using the same stars and symbols as Figure \ref{R21}. 
Stars with $R_{21}=0$ have undefined phases, and so do not appear in the
plot.}

\figcaption[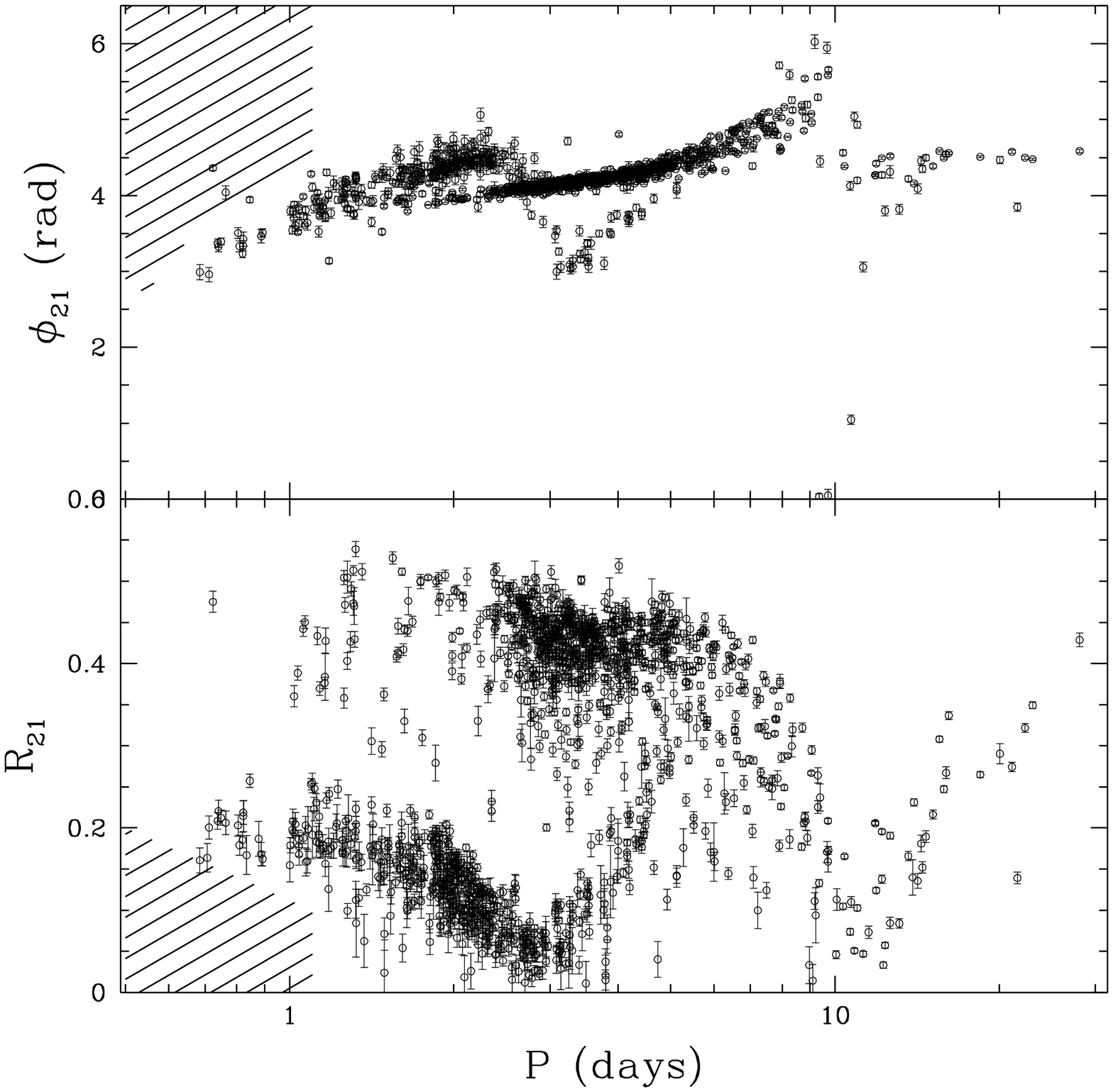]{
\label{R21-P21-for-LMC-Cephs}
The Fourier parameters $R_{21}$ and $\phi_{21}$ for MACHO 
Project Cepheids in the LMC, adapted from Welch \etal\ (1997).  The 
sequences in $R_{21}$ and $\phi_{21}$ extending to the longest periods 
are the \f\ mode sequences; the other sequences belong to the \fo\ mode.  
The hatched region delimits \so\ mode boundaries as found in Figures 
\ref{R21} and \ref{P21} for \so\ mode Cepheids with $R_{21}\neq 0$.  Any 
stars with $\phi_{21}$ in the hatched region have $R_{21}$ values 
indicative of \f\ and \fo\ pulsators.} 

\figcaption[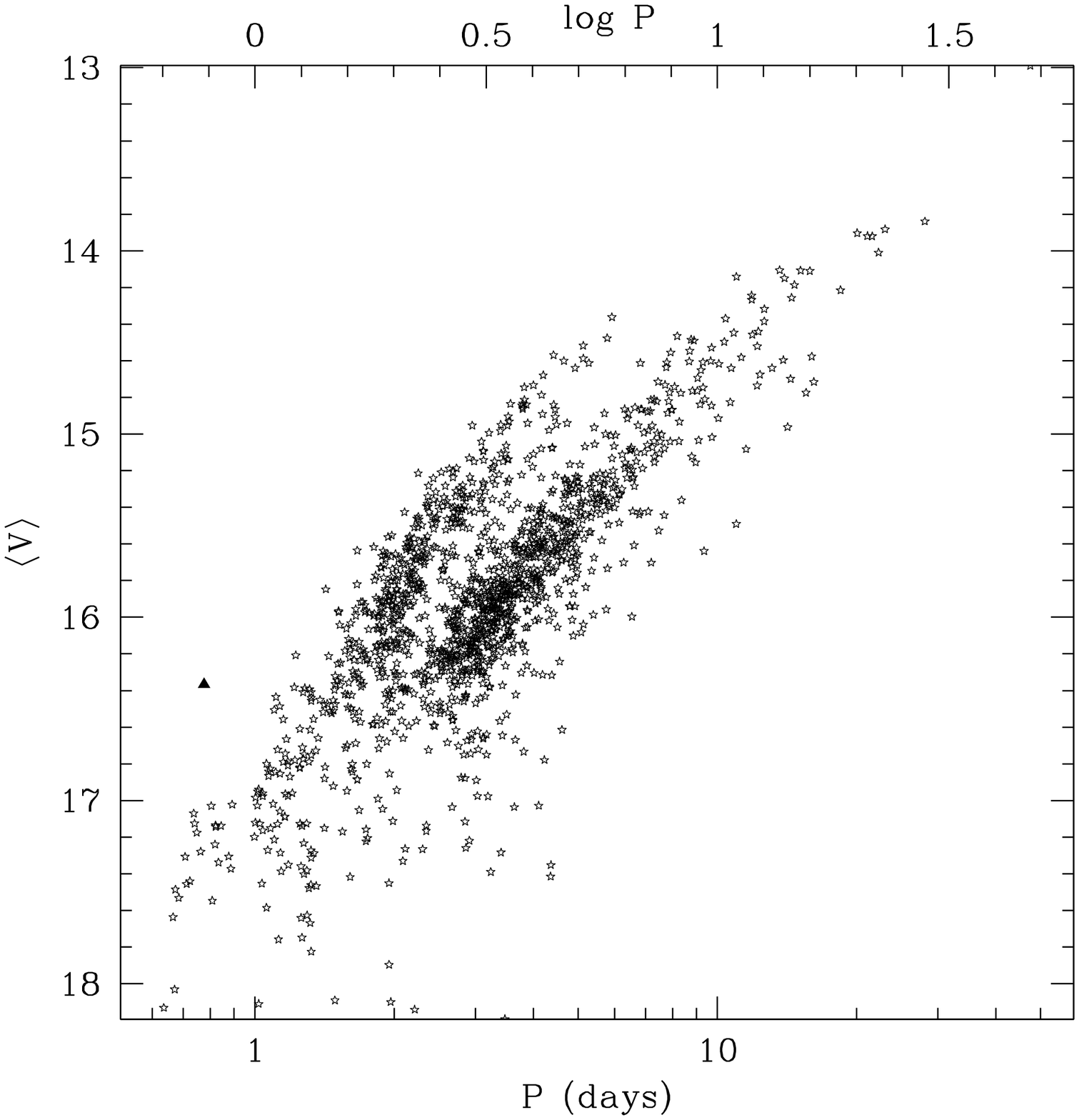]{
\label{PL}
The P--L relation for MACHO Project LMC Cepheids shown in Figure 
\ref{R21-P21-for-LMC-Cephs}, along with \socan, our \so\ candidate (the 
filled triangle).} 

\figcaption[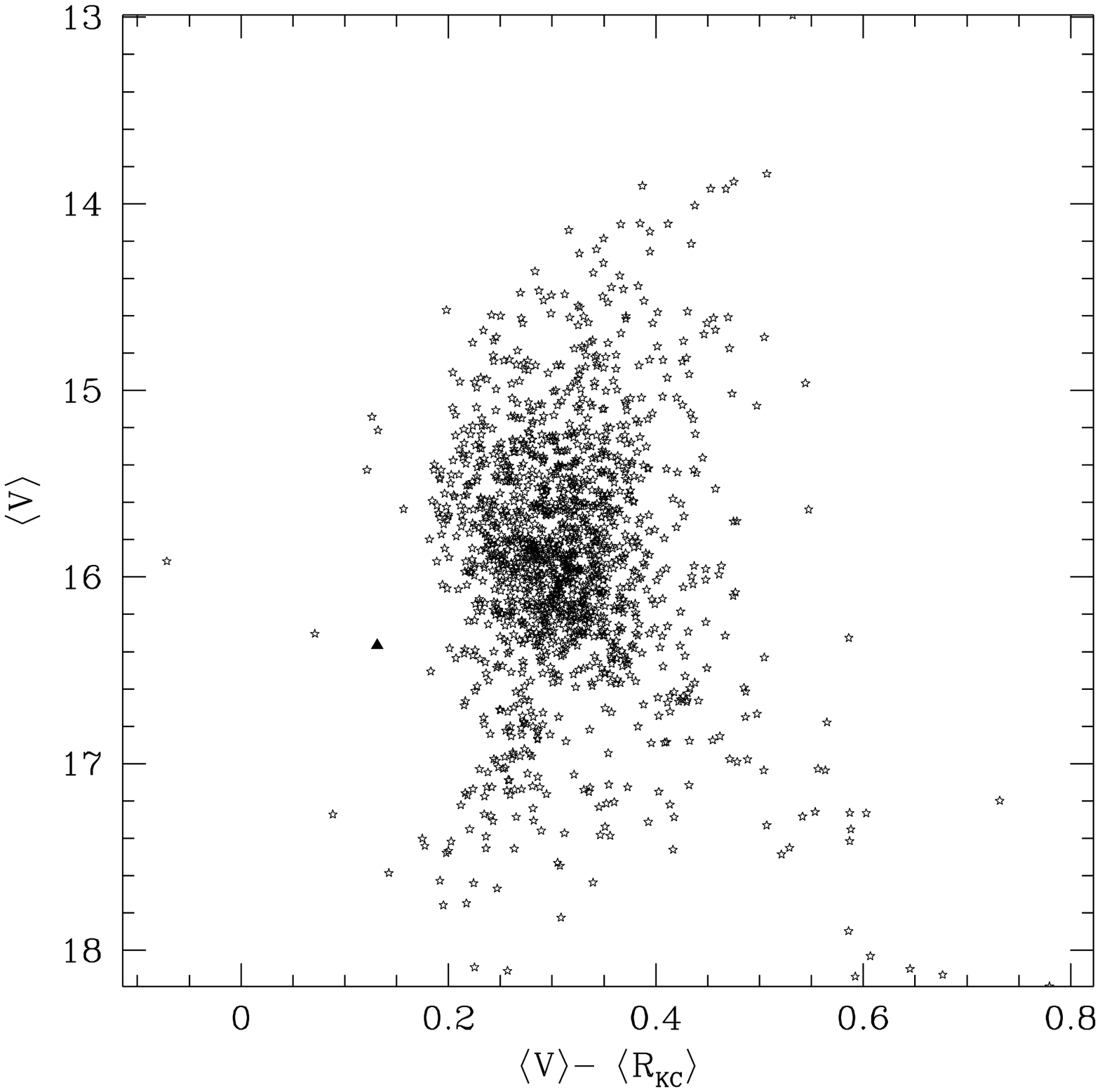]{
\label{CMD}
The CMD for MACHO Project LMC Cepheids from Figure 
\ref{R21-P21-for-LMC-Cephs}, along with \socan, our \so\ candidate (the 
filled triangle).} 

\figcaption[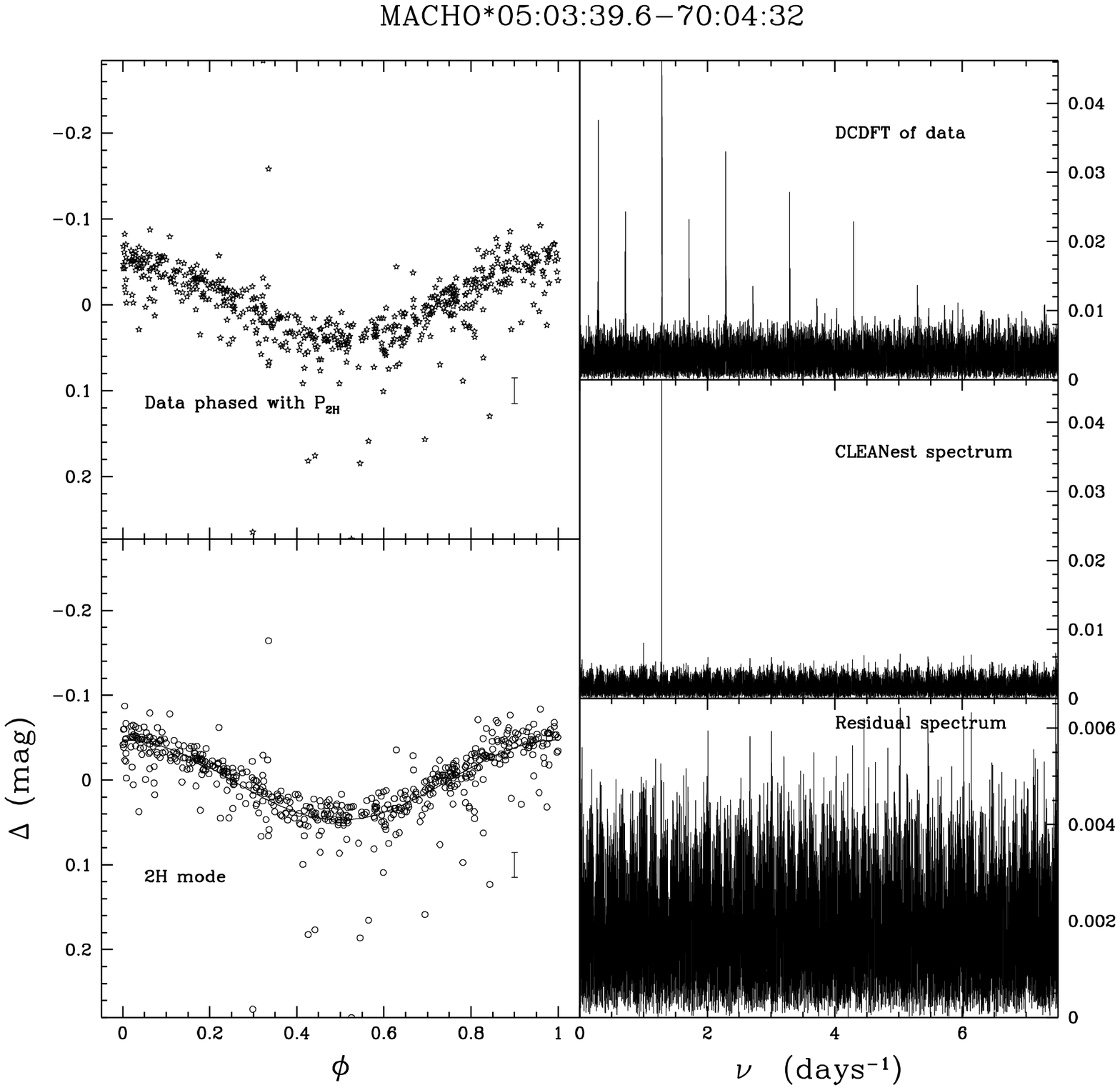]{
\label{V-fit}
Our photometry and model fit for \socan\ in the V band.  The 
left panels show the actual data and the model fit, phased with a period 
of 0.775961(19) days.  The lower panel does not include the sidereal day 
model term, but shows no appreciable difference from the top panel.  The 
right panels show the DCDFT of the raw data (top), the final residual 
DCDFT plus the best-fit frequencies ($\nu_\so$ and a sidereal frequency; 
middle), and the residual DCDFT (bottom).} 

\newpage

\begin{deluxetable}{rrrrr}
\footnotesize
\tablecaption{ LMC \fo/\so\ Beat Cepheid Periods and Period 
Ratios\label{pertable}} \tablewidth{0pt}

\tablehead{
\colhead{Cepheid\tablenotemark{a}} & 
\colhead{ID\tablenotemark{b}} & 
\colhead{$P_\fo$} &
\colhead{$P_\so$} &
\colhead{$P_\so/P_\fo$}\nl
\colhead{~} &
\colhead{~} &
\colhead{(days)} &
\colhead{(days)} &
\colhead{~}}

\startdata
{\tt MACHO*05:11:39.9$-$68:49:58} & {\tt 79..5022..339 }&{\tt 0.526264(09)} &{\tt  0.422854(12)} &{\tt 0.803502(26)} \nl
{\tt MACHO*04:53:15.5$-$68:16:37} & {\tt 47..2127..102 }&{\tt 0.578469(20)} &{\tt  0.466607(35)} &{\tt 0.806623(67)} \nl
{\tt MACHO*05:38:51.1$-$69:49:24} & {\tt 81..9485...45 }&{\tt 0.610981(13)} &{\tt  0.491236(16)} &{\tt 0.804012(31)} \nl
{\tt MACHO*05:48:55.2$-$70:30:03} & {\tt 12.11168..150 }&{\tt 0.640684(13)} &{\tt  0.516399(11)} &{\tt 0.806012(24)} \nl
{\tt MACHO*05:15:35.2$-$68:57:08} & {\tt 79..5747..197 }&{\tt 0.665188(20)} &{\tt  0.536387(11)} &{\tt 0.806369(30)} \nl
{\tt MACHO*05:15:05.4$-$69:39:55} & {\tt  5..5615.2564 }&{\tt 0.689983(15)} &{\tt  0.555710(12)} &{\tt 0.805398(25)} \nl
{\tt MACHO*05:15:06.3$-$69:39:54} & {\tt 78..5615...82 }&{\tt 0.689986(16)} &{\tt  0.555687(15)} &{\tt 0.805360(29)} \nl
{\tt MACHO*05:35:11.1$-$70:17:06} & {\tt 11..8873..189 }&{\tt 0.704752(16)} &{\tt  0.568415(13)} &{\tt 0.806547(26)} \nl
{\tt MACHO*05:16:28.6$-$69:36:34} & {\tt 78..5858..301 }&{\tt 0.711555(16)} &{\tt  0.574158(13)} &{\tt 0.806906(26)} \nl
{\tt MACHO*05:24:13.9$-$68:49:35} & {\tt 80..7080.2618 }&{\tt 0.715871(18)} &{\tt  0.577964(14)} &{\tt 0.807358(28)} \nl
{\tt MACHO*05:30:01.9$-$69:11:39} & {\tt 82..8042..142 }&{\tt 0.724083(18)} &{\tt  0.584623(15)} &{\tt 0.807399(29)} \nl
{\tt MACHO*05:26:02.1$-$69:52:10} & {\tt 77..7427..306 }&{\tt 0.729248(17)} &{\tt  0.588430(14)} &{\tt 0.806900(27)} \nl
{\tt MACHO*05:37:47.1$-$70:51:44} & {\tt 11..9348...78 }&{\tt 0.737788(18)} &{\tt  0.594408(14)} &{\tt 0.805662(27)} \nl
{\tt MACHO*05:26:01.5$-$69:30:42} & {\tt 77..7432..248 }&{\tt 0.740806(18)} &{\tt  0.595110(14)} &{\tt 0.803327(28)} \nl
{\tt MACHO*04:58:53.0$-$68:51:08} & {\tt 18..2965..104 }&{\tt 0.742537(18)} &{\tt  0.598306(14)} &{\tt 0.805759(27)} \nl
{\tt MACHO*05:16:28.5$-$69:25:36} & {\tt 79..5861.5053 }&{\tt 0.747881(18)} &{\tt  0.601368(14)} &{\tt 0.804097(27)} \nl
{\tt MACHO*05:16:28.5$-$69:25:36} & {\tt 78..5861..239 }&{\tt 0.747873(18)} &{\tt  0.601366(14)} &{\tt 0.804102(27)} \nl
{\tt MACHO*05:34:31.9$-$69:45:15} & {\tt 81..8760..204 }&{\tt 0.749809(19)} &{\tt  0.603618(16)} &{\tt 0.805029(29)} \nl
{\tt MACHO*05:15:49.4$-$68:41:53} & {\tt  2..5750.2010 }&{\tt 0.765468(19)} &{\tt  0.617622(15)} &{\tt 0.806856(28)} \nl
{\tt MACHO*05:09:29.8$-$68:21:20} & {\tt  2..4788...82 }&{\tt 0.768298(19)} &{\tt  0.620344(15)} &{\tt 0.807426(28)} \nl
{\tt MACHO*05:23:13.8$-$69:36:36} & {\tt 78..6947.2839 }&{\tt 0.783374(20)} &{\tt  0.631034(16)} &{\tt 0.805534(29)} \nl
{\tt MACHO*05:34:28.5$-$68:57:01} & {\tt 82..8772...88 }&{\tt 0.791231(21)} &{\tt  0.637010(17)} &{\tt 0.805088(31)} \nl
{\tt MACHO*05:38:45.4$-$70:36:12} & {\tt 11..9473..117 }&{\tt 0.807845(21)} &{\tt  0.650272(17)} &{\tt 0.804947(30)} \nl
{\tt MACHO*05:21:25.4$-$69:52:52} & {\tt 78..6701..236 }&{\tt 0.809502(21)} &{\tt  0.652649(17)} &{\tt 0.806236(29)} \nl
{\tt MACHO*05:23:59.1$-$69:15:31} & {\tt 80..7073..142 }&{\tt 0.834865(24)} &{\tt  0.673340(19)} &{\tt 0.806526(33)} \nl
{\tt MACHO*05:34:34.6$-$70:18:21} & {\tt 11..8751..129 }&{\tt 0.856625(24)} &{\tt  0.689217(19)} &{\tt 0.804573(31)} \nl
{\tt MACHO*05:46:42.8$-$70:40:50} & {\tt 12.10803..112 }&{\tt 0.859776(24)} &{\tt  0.692219(39)} &{\tt 0.805116(50)} \nl
{\tt MACHO*05:24:33.2$-$70:09:31} & {\tt  7..7181.1511 }&{\tt 0.889705(27)} &{\tt  0.715755(21)} &{\tt 0.804485(34)} \nl
{\tt MACHO*05:21:16.6$-$69:52:01} & {\tt 78..6580..150 }&{\tt 0.896293(39)} &{\tt  0.722064(26)} &{\tt 0.805612(46)} \nl
{\tt MACHO*05:23:10.0$-$70:28:45} & {\tt  6..6934...67 }&{\tt 0.920143(27)} &{\tt  0.740074(22)} &{\tt 0.804303(34)} \nl
{\tt MACHO*05:30:11.7$-$69:52:02} & {\tt 77..8032..175 }&{\tt 0.932517(28)} &{\tt  0.751390(28)} &{\tt 0.805765(38)} \nl
{\tt MACHO*05:49:27.6$-$71:32:07} & {\tt 15.11153...34 }&{\tt 0.963677(36)} &{\tt  0.775241(25)} &{\tt 0.804462(40)} \nl
{\tt MACHO*05:47:11.7$-$70:41:11} & {\tt 12.10803...77 }&{\tt 1.015125(33)} &{\tt  0.814284(47)} &{\tt 0.802151(53)} \nl
{\tt MACHO*05:07:44.6$-$68:35:20} & {\tt 19..4421..403 }&{\tt 1.017549(33)} &{\tt  0.816983(27)} &{\tt 0.802893(37)} \nl
{\tt MACHO*05:21:05.4$-$68:23:36} & {\tt  3..6602...41 }&{\tt 1.081246(40)} &{\tt  0.867335(32)} &{\tt 0.802163(42)} \nl
{\tt MACHO*05:10:15.3$-$68:20:28} & {\tt  2..4909...67 }&{\tt 1.084094(38)} &{\tt  0.870033(30)} &{\tt 0.802543(40)} \nl
{\tt MACHO*05:25:59.2$-$69:49:14} & {\tt 77..7428..149 }&{\tt 1.108837(39)} &{\tt  0.895047(58)} &{\tt 0.807194(60)} \nl
{\tt MACHO*05:45:22.1$-$70:50:13} & {\tt 12.10558..923 }&{\tt 1.119632(40)} &{\tt  0.896647(08)} &{\tt 0.800841(30)} \nl
{\tt MACHO*05:43:20.9$-$71:08:49} & {\tt 15.10191...50 }&{\tt 1.183158(45)} &{\tt  0.946917(36)} &{\tt 0.800330(43)} \nl
{\tt MACHO*05:09:08.0$-$68:56:43} & {\tt  1..4658...66 }&{\tt 1.217536(47)} &{\tt  0.978425(38)} &{\tt 0.803611(44)} \nl
{\tt MACHO*05:07:37.0$-$69:12:47} & {\tt  1..4412..130 }&{\tt 1.217973(48)} &{\tt  0.975255(38)} &{\tt 0.800719(44)} \nl
{\tt MACHO*05:49:28.9$-$70:22:40} & {\tt 12.11170...25 }&{\tt 1.251635(50)} &{\tt  1.008617(41)} &{\tt 0.805839(46)} \nl
{\tt MACHO*05:02:09.7$-$68:51:32} & {\tt  1..3570...55 }&{\tt 1.321184(56)} &{\tt  1.059515(45)} &{\tt 0.801944(48)} \nl
{\tt MACHO*05:20:19.7$-$70:42:29} & {\tt 13..6446...38 }&{\tt 1.341432(57)} &{\tt  1.074945(46)} &{\tt 0.801342(48)} \nl
{\tt MACHO*05:24:07.4$-$68:51:15} & {\tt 80..7079...62 }&{\tt 1.347889(58)} &{\tt  1.076446(46)} &{\tt 0.798616(49)} \nl
{\tt MACHO*05:37:36.2$-$69:44:21} & {\tt 81..9244...71 }&{\tt 1.352933(63)} &{\tt  1.081235(50)} &{\tt 0.799179(52)} \nl
{\tt MACHO*04:54:03.4$-$68:52:02} & {\tt 18..2239...43 }&{\tt 1.364210(59)} &{\tt  1.093286(48)} &{\tt 0.801406(49)} \nl
\enddata
\tablenotetext{a}{Object designations follow MACHO Project convention: {\tt 
MACHO*}, followed by an object's right ascension and declination (J2000.0).} 
\tablenotetext{b}{This column contains object designations internal to 
the MACHO Project.}
\end{deluxetable}

\newpage

\begin{deluxetable}{lrrrr}
\tablecaption{Fit Coefficients for \socan\ in Different Bandpasses\label{coeffs}} 
\tablewidth{0pt}

\tablehead{
\colhead{Band, $X$} &
\colhead{$X_0$ (mag)} &  
\colhead{$\Delta_X$ (mag)} &
\colhead{$\phi_X$ (rad)} &
\colhead{$P_X$ (days)}
}
 
\startdata

V            & {\tt 16.375$\pm$0.030} & {\tt 0.0472$\pm$0.0086} & {\tt 3.142$\pm$0.034} & {\tt 0.775961$\pm$0.000019}\nl
R$_{\rm KC}$ & {\tt 16.243$\pm$0.030} & {\tt 0.0369$\pm$0.0008} & {\tt 3.124$\pm$0.043} & {\tt 0.775961$\pm$0.000019}\nl
\enddata
\end{deluxetable}

\end{document}